\documentclass[10pt,a4paper]{article}
\pdfoutput=1
\usepackage{enumerate}
\usepackage{jheppub,undertilde}
\usepackage{amsthm}
\usepackage{graphicx}
\usepackage{siunitx}
\usepackage{diagbox}
\usepackage{textcomp,float,gensymb,wrapfig, enumitem,comment,dsfont,subfigure,framed,slashed,appendix}
\usepackage{hyperref}

\usepackage{soul}

\newcommand{\met}{E_T^{\rm miss}}

\preprint{SISSA 07/2018/FISI\\}

\title{LHC Phenomenology of Dark Matter with a Color-Octet Partner}

\author[a, b]{Alessandro Davoli}
\author[a, b]{Andrea De Simone}
\author[a, b]{Thomas Jacques}
\author[a]{Alessandro Morandini}

\affiliation[a]{SISSA, via Bonomea 265, 34136 Trieste, Italy}
\affiliation[b]{INFN Sezione di Trieste, via Bonomea 265, 34136 Trieste, Italy}

\emailAdd{alessandro.davoli@sissa.it}
\emailAdd{andrea.desimone@sissa.it}
\emailAdd{thomas.jacques@sissa.it}
\emailAdd{alessandro.morandini@sissa.it}

\abstract{
Colored dark sectors where the dark matter particle is accompanied by colored partners have recently attracted theoretical and phenomenological interest.
We explore the possibility that the dark sector consists of the dark matter particle and a color-octet partner, where the interaction with the Standard Model is governed by an effective operator involving gluons.
The resulting interactions resemble the color analogues of electric and magnetic dipole moments.
Although many phenomenological features of this kind of model only depend on the group representation of the partner under SU(3)$_c$, we point out that interesting collider signatures such as $R$-hadrons are indeed controlled by the interaction operator between the dark and visible sector.
We perform a study of the current constraints and future reach of LHC searches, where the complementarity between different possible signals is highlighted and exploited.
}
\arxivnumber{1803.02861}

\begin{document} 
\maketitle
\flushbottom

\section{Introduction}
It is well accepted that more than 80\% of the matter content of the universe is in the form of invisible dark matter (DM) \cite{Ade:2015xua}. 

Although its particle physics nature remains unknown, some properties can be inferred from experiments: for example, we know DM has to be (almost) electrically neutral, colorless and stable (at least on cosmological timescales). The search for DM proceeds primarily on three fronts: direct detection experiments look for the recoil of nuclei after interaction with DM~\cite{Akerib:2016vxi,Aprile:2017iyp,Cui:2017nnn}; indirect detection searches seek Standard Model (SM) particles resulting from DM annihilation~\cite{Adriani:2008zr,TheFermi-LAT:2017vmf,Aguilar:2013qda}; finally, collider experiments aim at producing DM  from SM states~\cite{Basalaev:2017hni,Aaboud:2017phn,Sirunyan:2018gka}.

Of course, DM could just be the lightest state of a whole dark sector, consisting of several other particles, which may carry electric or color charges. 
Particularly interesting is the case in which the DM is accompained by a long-lived particle (LLP) which travels a measurable distance before decaying~\cite{Kaplan:2009ag,Baumgart:2009tn,Dienes:2012yz,Kim:2013ivd,Co:2015pka,Hochberg:2015vrg,Davoli:2017swj}. LLPs have also been studied in different contexts of physics beyond the Standard Model (BSM), e.g. supersymmetry or composite-Higgs models \cite{Wells:2003tf,ArkaniHamed:2004fb,Arvanitaki:2012ps,ArkaniHamed:2012gw,Cui:2014twa,Csaki:2015uza,Barnard:2015rba}.

The possibility that the dark sector consists of colored particles in addition to the dark matter has also attracted recent interest \cite{deSimone:2014pda, Baker:2015qna, ElHedri:2017nny, Garny:2017rxs}.
In the context of LHC searches for the dark matter, this scenario is particularly remarkable because the phenomenology benefits from enhanced QCD-driven production rates of the colored partners.

A great deal of phenomenological properties of colored dark sectors are somewhat model-independent, in the sense that they only depend on the representation of the colored partner under the SU(3)$_c$ gauge group \cite{deSimone:2014pda}.
However, as we will discuss below, important features for collider phenomenology are indeed reliant on the interaction between the dark sector and the SM.

In this paper, we consider the SM augmented by a dark sector  constisting of a DM particle  and a nearly-degenerate colored state, in the adjoint representation of SU(3)$_c$.
This dark sector communicates with the SM via 
 a dimension-5 effective operator (the validity of effective theories for DM searches has been widely discussed in the literature, see e.g. Refs.~\cite{Busoni:2013lha,Morgante:2014kra,Busoni:2014uca,Busoni:2014haa,Busoni:2014sya,Bell:2015sza,DeSimone:2016fbz,Bruggisser:2016nzw,Kahlhoefer:2017dnp}). 
Such a scenario is particularly interesting because the colored partner could hadronize in bound states like ordinary quarks and gluons. In a supersymmetric context this is a well-known possibility, and such bound states, originally introduced in Ref.~\cite{Farrar:1978xj}, are called $R$-hadrons. We use here the same terminology, although our considerations do not assume any underlying supersymmetry. For more recent papers about $R$-hadrons, see e.g. Refs.~\cite{Aaboud:2016uth,Aaboud:2016dgf,CMS:2017rlw,Bond:2017wut,Garny:2018icg}.
In addition, since the decay of the colored partner is governed by a suppressed non-renormalizable operator, such a bound state can easily travel macroscopic distances and leave tracks in the collider detector.

The paper is organized as follows: in section~\ref{sec:model}, we introduce the model and discuss some of its features and implications; in section~\ref{sec:analysis}, we consider LHC constraints derived from monojet and $R$-hadron searches, focusing on the interplay between them. Finally, we conclude in section~\ref{sec:conclusion}.

\section{Chromo-electric dipole dark matter}
\label{sec:model}

\subsection{Model}
\label{subsec:model}

Dark matter, despite being neutral, can be coupled to colored Standard Model particles. In order to allow such a coupling, colored particles within the dark sector are required~\cite{Pierce:2007ut,Hamaguchi:2014pja,Ibarra:2015nca,Ellis:2015vaa,Liew:2016hqo,Mitridate:2017izz,Garny:2017rxs,DeLuca:2018mzn,Biondini:2018pwp}. 
In this work, we consider an extension of the minimal scenario, where the DM particle $\chi_1$ is accompanied by a slightly heavier
partner $\chi_2$. We denote the masses of these particles by $m_1$ and $m_2\equiv m_1+\Delta m$, respectively. Both $\chi_1$ and $\chi_2$ are Majorana fermions.

At the renormalizable level, scalar or fermionic partners in the fundamental representation of SU(3) can be responsible for the coupling of DM with the SM quarks~\cite{Garny:2014waa,Giacchino:2015hvk,Garny:2017rxs}.
If, instead, we are to consider a coupling to gluons, the lowest dimensional operator has $D=5$ and involves a colored partner $\chi_2$ in the adjoint representation of SU(3).
If we denote the dark matter particle by $\chi_1$, the free Lagrangian for the dark sector is:
\begin{equation}
\mathcal{L}_0=\frac{1}{2}\bar{\chi}_1\left(i\slashed{\partial}-m_1\right)\chi_1+\frac{1}{2}\bar{\chi}_2^a\left(i\slashed D-m_2\right)\chi_2^a\,,
\end{equation}
with $a$ being the color index in the adjoint representation.

The coupling to gluons can be attained via effective operators mimicking the (chromo-) electric and (chromo-)magnetic dipole moments, as follows:
\begin{equation}
\mathcal{L}_\mathrm{int}=\frac{i}{2 m_1}\bar{\chi}_2^a\sigma^{\mu\nu}\left(\mu_\chi-id_\chi\gamma^5\right)\chi_1 G^a_{\mu\nu}\,,
\label{chromo_int}
\end{equation}
where $\sigma^{\mu\nu}=i/2[\gamma^\mu,\gamma^\nu]$ and $G_{\mu\nu}^a$ is canonically normalized.

The two operators in Eq.~\eqref{chromo_int} give rise to similar phenomenology and no interference effect arises in any of the observables we study in this paper. Therefore, for simplicity, we limit ourselves to study only the operator with $d_\chi$ in the rest of the paper.

Effective operators describing dipole moments typically arise after integrating out heavy particles of the underlying ultraviolet theory at loop-level. If this is the case, the operator in Eq.~\eqref{chromo_int} should be further suppressed by $\alpha_s/(4\pi)$, and so the importance of higher-order operators may not be negligible. A more complete theoretical analysis of the origin of the interaction in Eq.~\eqref{chromo_int} and of the role of higher-order operators is left to a future work.

The interactions of the dark sector with the SM particles are then described by the parameters $\{m_1,\Delta m,d_\chi\}$. In particular, we require that $d_\chi \ll 1$: this interaction term, in fact, could be written as an effective term suppressed by $1/\Lambda$, with $\Lambda$ being the scale of some underlying new physics. It is then natural to formally identify $d_\chi\sim m_1/\Lambda$, which has to be small in order for the effective theory to be reliable. 
For the values of $d_\chi$ considered in our analysis,  the energy scales of the processes of interest are always well below the operator scale $\Lambda$, thus ensuring we are in the regime of valid effective field theory.

The simplest process leading to the decay of $\chi_2$ is $\chi_2\rightarrow \chi_1 g$, whose width, at leading order in $\Delta m/m_1$, is:
\begin{equation}
\label{eq:x2decay}
\Gamma_{\chi_{2}}=\frac{d_\chi^2}{\pi}\frac{\Delta m^3}{m_1^2}.
\end{equation}
Since $d_\chi$ is required to be small, we would naturally expect $\chi_2$ to be a long-lived particle with lifetime on the detector timescale, as will be explored later.

\subsection{Relic density}\label{sec:relic}

A first constraint on the parameter space can be obtained by requiring that the model reproduces the observed dark matter abundance $\Omega h^2=0.1194$ \cite{Ade:2015xua}. Such a relic density is determined by processes of the form $\sigma(\chi_i\chi_j\rightarrow \text{SM}\,\text{SM})$. The expressions for the corresponding cross sections, at leading order in $\Delta m/m_1$ and $m_f/m_1$ (where $m_f$ is the mass of a generic SM fermion $f$), are shown in table~\ref{table_sigma_chromo}, although the complete expressions, which can be found in Appendix \ref{app:sigma}, have been used in the calculations. In order to determine the relic density predicted by this model, two modifications to the standard procedure have, in principle, to be taken into account: first, if the mass splitting between $\chi_1$ and $\chi_2$ is small compared to their masses, co-annihilations must be included~\cite{Griest:1990kh,Baker:2015qna}; second, due to the color charge of $\chi_2$, Sommerfeld enhancement (introduced below) modifies the value of $\sigma(\chi_2\chi_2\rightarrow \text{SM}\,\text{SM})$.

\begin{table}[t]
\begin{center}
\begin{tabular}{|c||c|c|}\hline
\diagbox[dir=NW,width=2.6cm, height=1.2cm]{\;\;\;\;\;\;$ij$}{$SM SM$} & $q\,\bar q$ & $g\,g$\\\hline\hline
 & $\dfrac{}{}$& \\[-1em]
11 &  & $\dfrac{2\,d_\chi^4}{\pi}\,\dfrac{1}{m_1^2}\,$ \\
 & & \\[-1em]\hline
 & & \\[-1em]
$12$ & $\dfrac{d_\chi^2\,g_s^2}{96\pi}\,\dfrac{1}{m_1^2}v^2$ & $\dfrac{3\,d_\chi^2\,g_s^2}{16\pi}\,\dfrac{1}{m_1^2}$ \\
 & & \\[-1em]\hline
 & & \\[-1em]
 $22$ &$\dfrac{3\,g_s^4}{256\pi}\,\dfrac{1}{m_1^2}$ & $\dfrac{27\,g_s^4}{512\pi}\,\dfrac{1}{m_1^2}\,+\, \mathcal O(d_\chi^2)$
 \\[1em]\hline
\end{tabular}
\end{center}
\caption{\label{table_sigma_chromo}Different contributions to the effective cross-section $\langle\sigma v\rangle_{\chi_i\chi_j\to SM SM}$. The QCD coupling is denoted by $g_s$, while $v$ is the relative velocity in the $\chi_i\chi_j$ center-of-mass frame.}
\end{table}

As far as the co-annihilations are concerned, the effective cross-section which determines the observed abundance of DM in the universe is:
\begin{equation}
{\langle\sigma v\rangle}_\mathrm{eff}=\frac{1}{{\left(1+\alpha\right)}^2}\left({\langle\sigma v\rangle}_{11}+2\alpha{\langle\sigma v\rangle}_{12}+\alpha^2{\langle\sigma v\rangle}_{22}\right)\,,
\label{sigma_eff}
\end{equation}
where $\alpha\equiv g_2/g_1{(1+\Delta m/m_1)}^{3/2}e^{-x\Delta m/m_1}$, $x\equiv m_1/T$, ${\langle\sigma v\rangle}_{ij}\equiv{\langle\sigma v\rangle}_{\chi_i\chi_j\to SM\,SM}$ and $g_i$ is the number of degrees of freedom of $\chi_i$.
The relic abundance is then related to this effective cross-section as:
\begin{equation}
\Omega h^2=\frac{0.03}{\displaystyle\int_{x_F}^\infty dx\,\frac{\sqrt{g_*}\,}{x^2}\,\frac{{\langle\sigma v\rangle}_\mathrm{eff}}{\SI{1}{\pico\barn}}}\,,
\label{relic_abundance}
\end{equation}
where $g_*$ is the number of relativistic degrees of freedom at the freeze-out temperature $T_F$, determined by the implicit equation:
\begin{equation}
x_F=25+\log\left[\frac{1.67\,n_F}{\sqrt{g_*x_F}}\,\frac{m_1}{\SI{100}{\giga\electronvolt}}\,\frac{{\langle\sigma v\rangle}_\mathrm{eff}}{\SI{1}{\pico\barn}}\right]\,,
\end{equation}
with $n_F= g_1(1+\alpha)$ being the effective number of degrees of freedom of the system $(\chi_1,\chi_2)$.
In the following, we take $g_*=106.75$ as a reference.

As already mentioned, Sommerfeld enhancement also plays an important role in the determination of the relic abundance~\cite{ANDP:ANDP19314030302,Feng:2010zp,ElHedri:2016onc}: Sommerfeld enhancement is a non-perturbative effect due to the exchange of soft gluons betweeen the colored particles in the initial state. This is therefore relevant for the self-annihilation of $\chi_2$. Model independent discussion of this effect can be found in Refs.~\cite{deSimone:2014pda,ElHedri:2017nny}. These analyses assume that the relic DM density is dominated by QCD, remaining agnostic about the particular phenomenology deriving from the new BSM coupling. This is a reasonable assumption for the model we consider since, as already stated, it is natural (and indeed necessary) to assume that $d_\chi\ll1$. Co-annihilations where the DM annihilation cross-section contributes negligibly to the relic density have been recently analyzed in Ref.~\cite{DAgnolo:2018wcn} in the more general context of \emph{sterile co-annihilations}.

When the final state is characterized by a single representation $Q$, the Sommerfeld-corrected cross-section is $\sigma_\text{Somm}=S\left(C_Q \alpha_s/\beta\right)\sigma_\text{Pert}$, where $S$ is the non-perturbative correction depending on the final representation (through the Casimir element $C_Q$) and on the velocity of the particles $\beta$. If, on the other hand, we have more than one possible final state representation, we need to consider the decomposition $R\otimes R'=\oplus_Q Q$, where $R$ and $R'$ are the initial state representations (in our case $R=R'=8$) and $Q$'s are the final state ones. Each representation $Q$ gives a contribution to the total cross-section and has its own value of $C_Q$. After group decomposition, the final result is given by eqs.~(2.24, 2.25) of Ref.~\cite{deSimone:2014pda}. 

\begin{figure}
\centering
\includegraphics[width=0.5\linewidth]{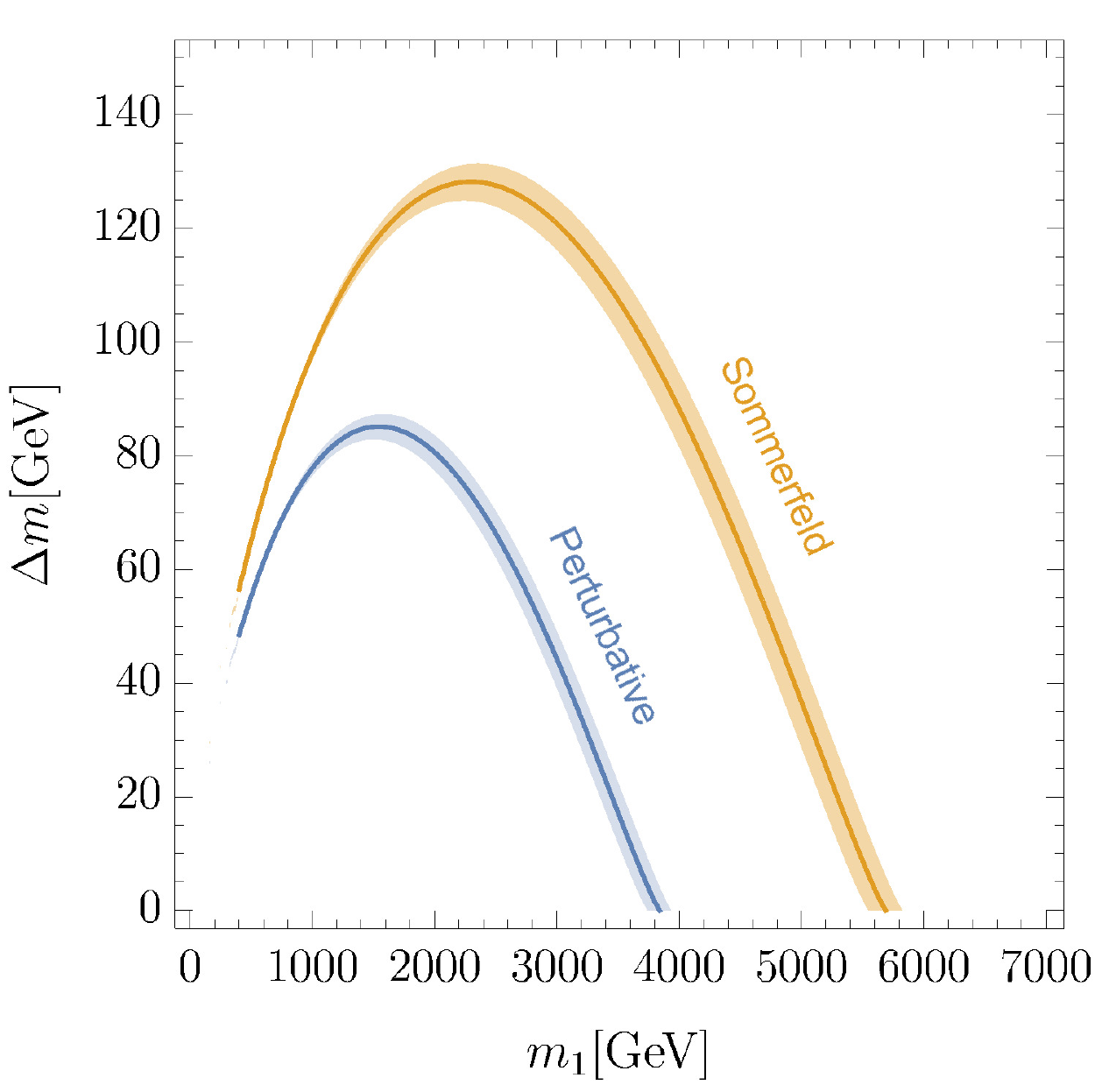}
\caption{\label{relic:chromo}
Contours corresponding to the measured relic abundance $\Omega h^2=0.1194\pm
0.0022\, (1\sigma)$, together with its 3-$\sigma$ bands, in the case of domination by QCD processes. The perturbative result and the result including Sommerfeld enhancement are both shown. Note that part of the parameter space is already excluded by LHC searches, as explained in section~\ref{sec:comparison} and shown in fig.~\ref{fig:chromo_comparison}.}
\end{figure}

As a result, the contour yielding the correct relic density, with and without the inclusion of such a non-perturbative effect, can be found in fig.~\ref{relic:chromo}. In this plot, we only consider the dominant contributions from QCD self-annihilations, not including the sub-leading contributions of processes proportional to $d_\chi$, which will be negligible if $d_\chi \ll 1$. This is actually well motivated from the previous discussion about the magnitude of $d_\chi$. 

\begin{figure}
\centering
\includegraphics[width=0.4\linewidth]{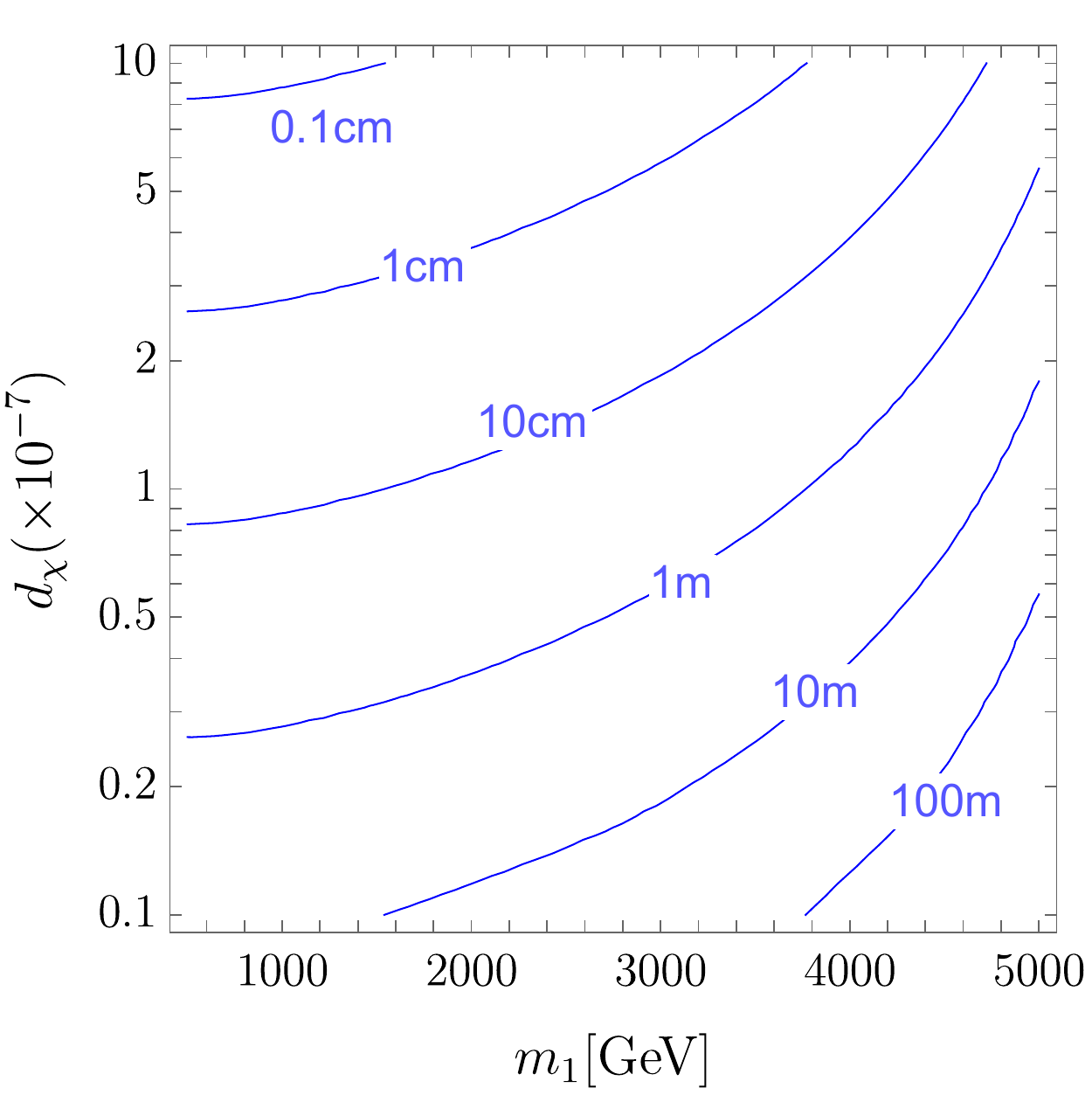}
\caption{Decay lengths at rest for the heavy partner $\chi_2$ in a parameter space where $\Gamma_{\chi_2}^{-1}$ is macroscopic (in the $\SI{}{\centi\meter}-\SI{}{\meter}$ range). The mass splitting $\Delta m$ is fixed, for given $m_1$, by the relic density as shown in fig.~\ref{relic:chromo}. Small values of $d_\chi$ and large values of $m_1$ give origin to larger values of decay length.  }\label{fig:chromo_decay}
\end{figure}

While the relic density is dominated by QCD processes, we see from eq.~\eqref{eq:x2decay} that the decay length instead depends quadratically on $d_\chi$.
Therefore the smallness of $d_\chi$ leads to macroscopic decay lengths, which are an interesting feature we will use in our analysis.
From here on, we fix  the mass splitting $\Delta m$ as a function of the mass of the DM candidate $m_1$,
using the Sommerfeld corrected curve in fig.~\ref{relic:chromo}. 
This imposes the correct relic density for all points in parameter space that  we consider. As a consequence, the decay length now only depends on the mass of $\chi_1$ and the coupling $d_\chi$. The full numerical results for the decay length can be found in fig.~\ref{fig:chromo_decay}, where we show contours of the proper decay length of $\chi_2$.

\subsection{Departure from chemical equilibrium}
The co-annihilation paradigm, described in section~\ref{sec:relic}, implicitly assumes that chemical equilibrium is maintained until the DM freeze-out.
Under particular conditions, however, it is possible that this assumption might not be valid~\cite{Garny:2017rxs}: this can happen, for instance, when the relic abundance is dominated by a SM (here, QCD) coupling: in this case, the coupling characterizing the BSM physics ($d_\chi$ in the case at hand) remains unconstrained. This translates into the fact that very small values for such a coupling are in principle allowed, leading to a possible breakdown of chemical equilibrium.

The important ratios to evaluate are $\Gamma_{\chi_i\chi_j}/H$, where $\Gamma_{\chi_i\chi_j}$ generically represents the rate of a process involving $\chi_i$ and $\chi_j$: it can be the scattering $\chi_2\,\chi_2\to\chi_1\,\chi_1$, the decay $\chi_2\to\chi_1\,g$, the conversions $\chi_2\,g\to\chi_1\,g$ and $\chi_2\,q\to\chi_1\,q$, as well as all the inverse reactions.

The rates of decay and conversion are proportional to $d_\chi^2$, while that for the scattering is proportional to $d_\chi^4$; therefore this latter process is expected to be the most sensitive to $d_\chi$, and is therefore expected to have the smallest rate.

\begin{figure}
\centering\includegraphics[width=0.9\textwidth]{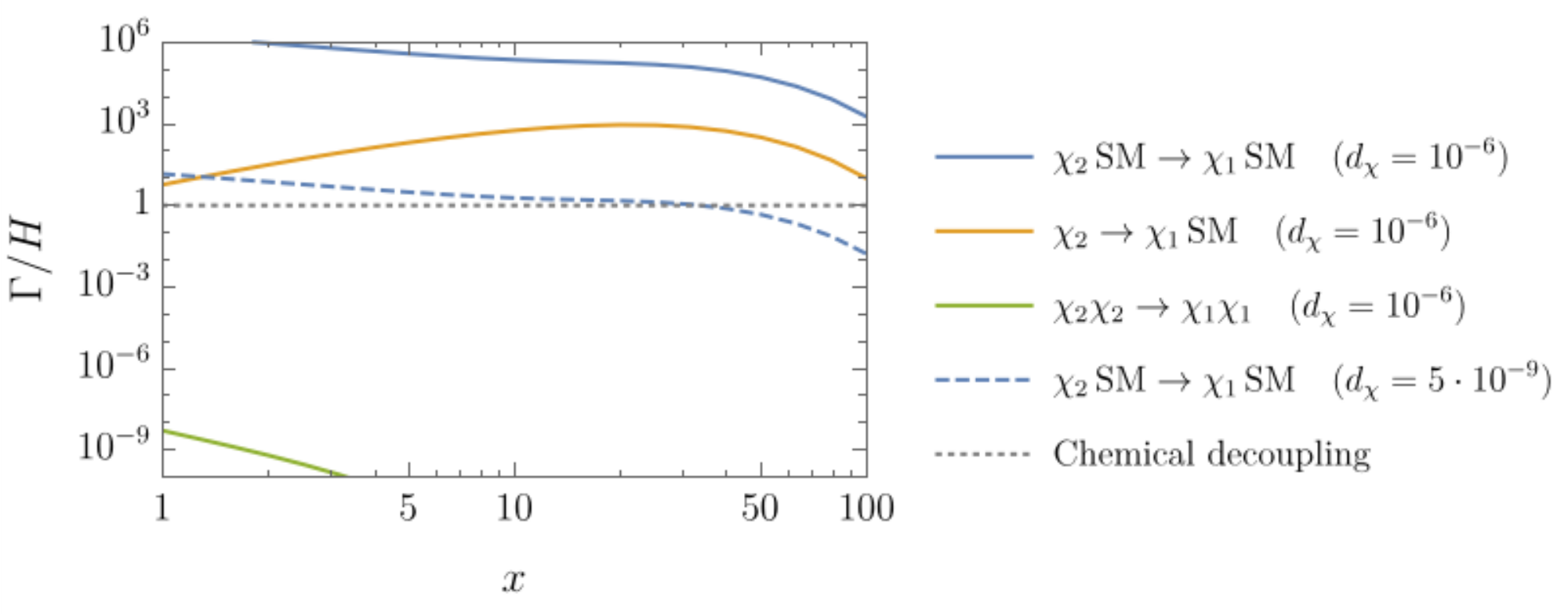}
\caption{Interaction rates for the case $m_1=\SI{1}{\tera\electronvolt}$ and $d_\chi=\SI{e-6}{}$. Different choices for the mass of the DM give similar results, and the scattering with gluons always turns out to be the relevant contribution for the determination of departure from chemical equilibrium.}\label{fig:chromo_chem}
\end{figure}

When the largest of these rates $\Gamma_{\chi_i\chi_j}^{\mathrm{(max)}}$ is such that $\Gamma_{\chi_i\chi_j}^{\mathrm{(max)}}/H\lesssim1$, the assumption of chemical equilibrium (which eq.~\eqref{sigma_eff} relies on) ceases to be valid. If this is the case, a numerical integration of the complete set of Boltzmann equations, including conversions, is necessary. The ratios $\Gamma/H$ for these three processes (in the direction $\chi_2\to\chi_1$) are shown in fig.~\ref{fig:chromo_chem}.

Since the rate corresponding to $\langle\sigma v\rangle_{\chi_2\,g\leftrightarrow\chi_1\,g}\propto g_s^2\,d_\chi^2$ turns out to be the dominant contribution, scatterings with gluons are ultimately responsible for maintaining chemical equilibrium.

In order to test the possible breakdown of chemical equilibrium before and during freeze-out (where eq.~\eqref{sigma_eff} has its validity), the ratio $\Gamma_{\chi_2\,\chi_1}/H\equiv H^{-1}\,n\langle\sigma v\rangle_{\chi_2\,g\leftrightarrow\chi_1\,g}$ can be investigated in the region $20\lesssim x_F\lesssim30$. From fig.~\ref{fig:chromo_chem}, we see that in this region $\Gamma_{\chi_2\,\chi_1}/H\sim10^{4}$ for $d_\chi=10^{-6}$. We can therefore estimate the breakdown of chemical equilibrium to occur when:
\begin{equation}
\frac{\Gamma_{\chi_2\,\chi_1}}{H}\lesssim1\quad\Leftrightarrow\quad d_\chi\lesssim10^{-8}\,.
\end{equation}
This simple scaling argument is actually in agreement with the explicit result shown in  fig.~\ref{fig:chromo_chem}.

In the following, we therefore assume $d_\chi\gtrsim10^{-8}$, in order to be in the regime of chemical equilibrium.

\section{LHC searches}\label{sec:analysis}

In this section we analyze the constraints on the model coming from the two most important channels: $R$-hadrons and monojet.
In principle, it would also be possible to have limits from dijet-resonance bounds coming from the production and fragmentation of a bound state of two $\chi_2$ particles, similar to a gluinonium. Since this results in rather weak constraints, we have described it in Appendix~\ref{app:dijet}.

\subsection{$R$-hadron constraints}\label{sec:rhadrons}

The color charge of the $\chi_2$ particle implies that it can hadronize with SM particles on the detector timescale, forming particles analogous to the $R$-hadrons in supersymmetry. If stable on a detector timescale, these colorless composite states can be detected via an ionization signature as they travel through the detector at speeds significantly less than the speed of light.

We apply ATLAS constraints on the $\chi_2$ production cross-section from Ref.~\cite{Aaboud:2016uth}, which searches for $R$-hadrons at $\sqrt{s} = 13$ TeV with 3.2 fb$^{-1}$ of data. The relevant constraints are those on gluinos, since $\chi_2$ is a color octet. 

We also consider an approximate high-luminosity (HL) projection of these limits to $\mathcal{L}=\SI{3000}{\per\femto\barn}$, using the procedure outlined in Ref.~\cite{Barnard:2015rba}, applied to the ATLAS analysis. The relevant results are the background counts in Table 3 of Ref.~\cite{Aaboud:2016uth} for the gluino search, which we  rescale with the increased luminosity. We assume that the same efficiencies of Table 3 apply to the HL bounds. It should be noted that in the HL regime the 
results are limited by systematics rather than statistics. 
The signal simulations are the same for the current luminosity and for higher luminosities.

In order to simulate the pair production of $\chi_2$ particles at parton level we have used \textsc{Madgraph5\_aMC@NLO}~\cite{Alwall:2014hca}, where the model has been implemented using \textsc{FeynRules} \cite{Alloul:2013bka}, and apply the $R$-hadronisation routine from \textsc{Pythia~8.230}~\cite{Sjostrand:2007gs}. The probability of each $\chi_2$ being stable at least up to the edge of the ATLAS calorimeter is given by
\begin{equation}
\mathcal{P}(\ell > \ell_{\rm calo}) = \exp\left(-\frac{\ell_{\rm calo}}{\ell_T}\right),
\end{equation}
where $\ell_{\rm calo}=\SI{3.6}{\meter}$ is the transverse distance to the edge of the calorimeter and we defined $\ell_T=p_2^T/(m_2\,\Gamma_{\chi_2})$. This probability is applied on an event-by-event basis to find the effective cross-section of events yielding at least one $R$-hadron. This relies on the assumption that the lifetime of the resultant $R$-hadron is at least as long as the unhadronized $\chi_2$ lifetime. Following Ref.~\cite{Aaboud:2016uth}, we assume that 90\% of the $\chi_2$ form charged $R$-hadrons. 

\begin{figure}[t]
\centering
\includegraphics[width = 0.525\linewidth]{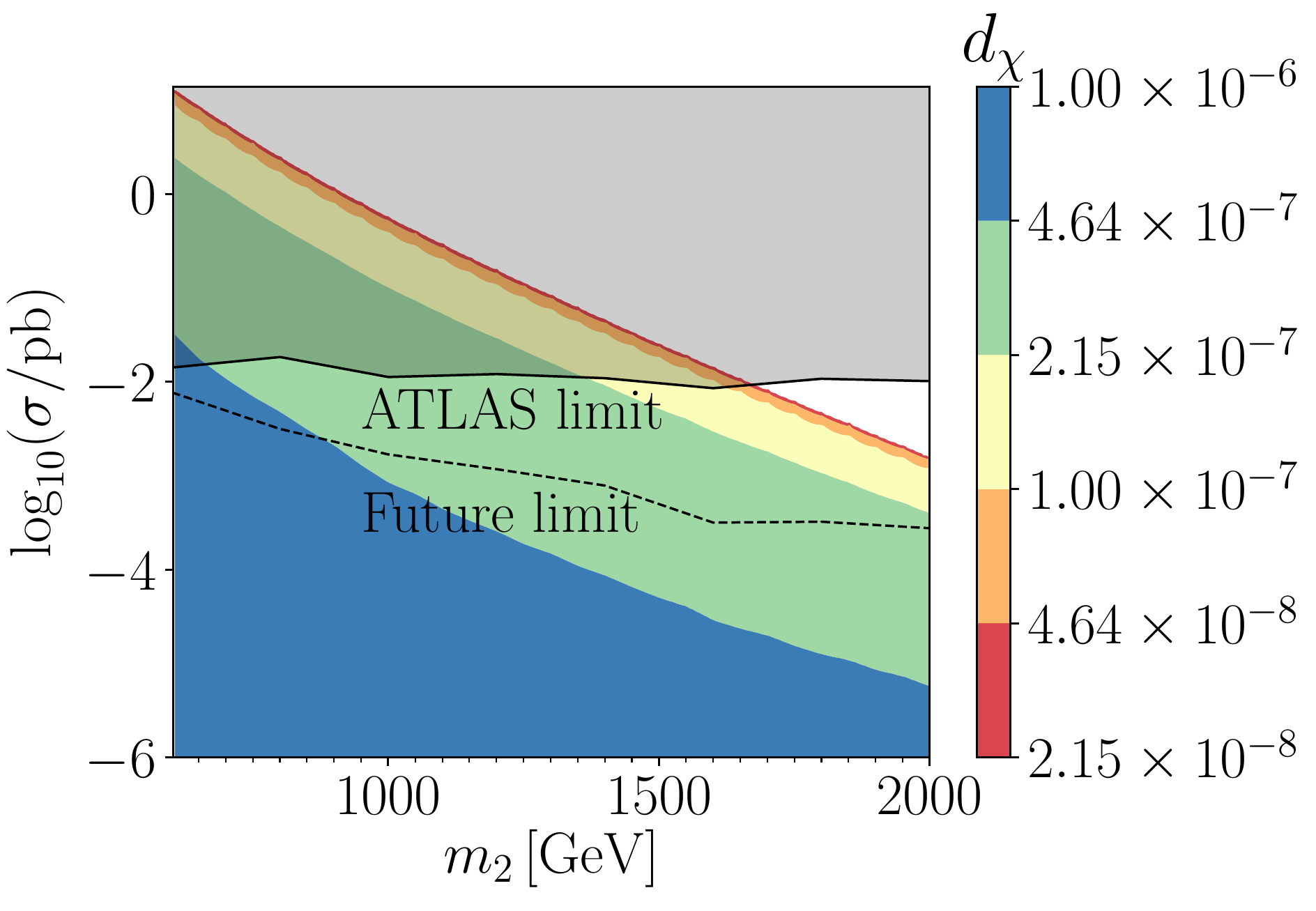}
\includegraphics[width = 0.375\linewidth]{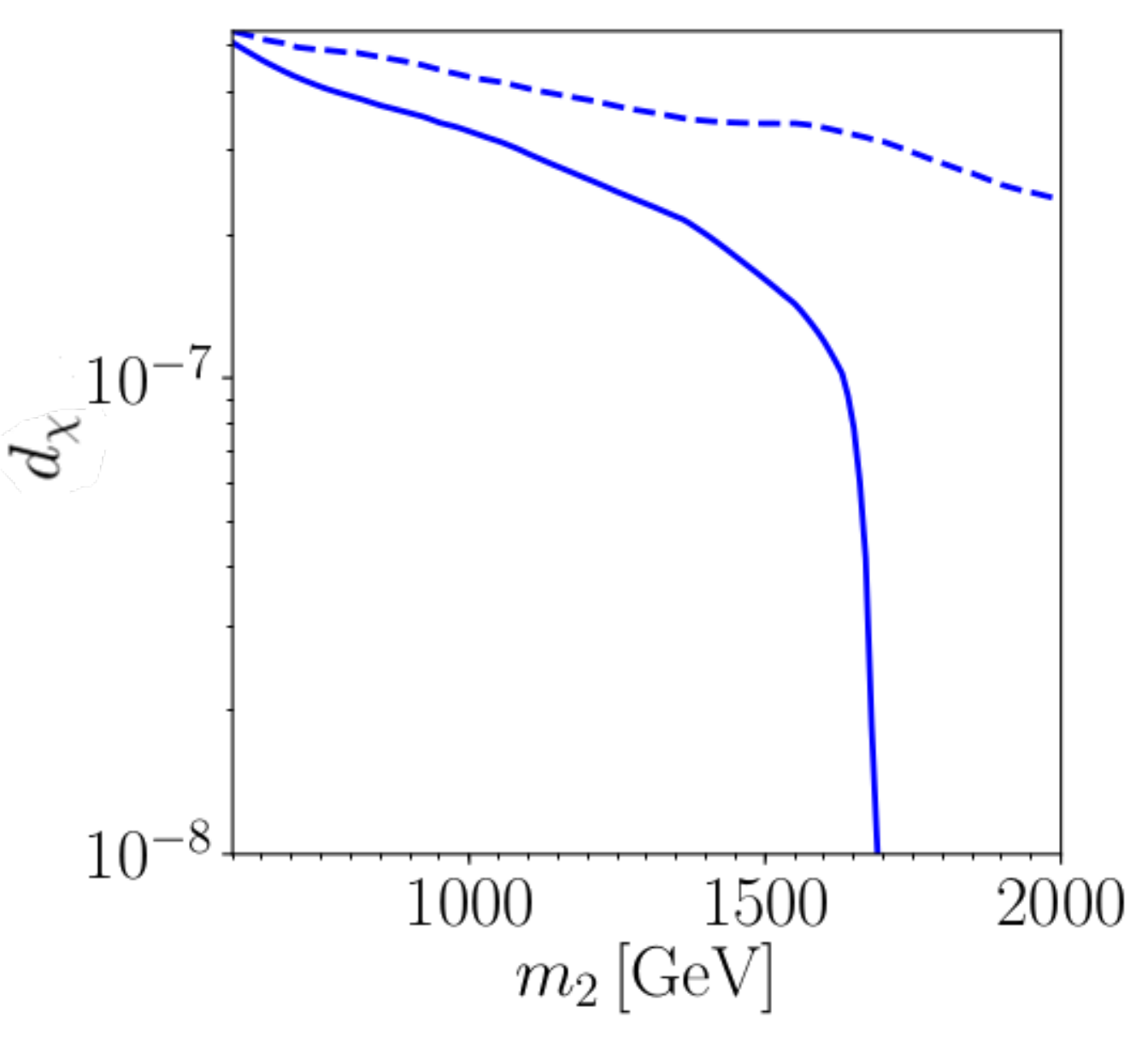}
\caption{\label{fig:r_hadrons}\emph{Left:} Contours showing the value of the coupling $d_\chi$ yielding a given production cross-section for calorimeter-stable $\chi_2$, overlayed with the region currently excluded by ATLAS and a high-luminosity projection of exclusions. \emph{Right:} Lower limit on $d_\chi$ from current (solid) and projected high-luminosity (dashed) $R$-hadron constraints as a function of the $\chi_2$ mass. Smaller values of $d_\chi$ will increase the $\chi_2$ decay length, exceeding the limit on the production cross-section of calorimeter-stable $\chi_2$. Note that $m_2$ is related to the DM mass by the values of $\Delta m$ given in fig.~\ref{relic:chromo}.}
\end{figure}

Contours showing the relationship between this effective production cross-section, $m_2$ and $d_\chi$ are shown in Fig.~\ref{fig:r_hadrons}, along with current and projected future ATLAS limits on the cross-section. 

\subsection{Monojet }\label{sec:monojet_general}

A generic particle physics model for DM is usually sensitive to so-called `monojet' searches, where  DM produced in a collider recoils from a high-energy jet, leaving a large missing energy ($\met$) signature as it passes through the detector without interacting~\cite{Aaboud:2016tnv,Vannerom:2016pxv,Aaboud:2017phn,Sirunyan:2017qaj}.

For the chromo-electric model the production processes leading to the monojet signature are of the form $p\,p\,\rightarrow\,\chi_i\,\chi_l\, j$ with $i, l \in \{1,2\}$. Since $d_\chi\ll 1$,  the leading contribution will be from the QCD-mediated production channel $p\,p\,\rightarrow\,\chi_2\,\chi_2\, j$, since all other terms are proportional to powers of $d_\chi$. In this regime, the relic density profile shown in fig.~\ref{relic:chromo} will apply.

We apply the latest monojet constraints from ATLAS~\cite{Aaboud:2017phn}, which searches for events with large missing energy and at least one high-energy jet, with center-of-mass energy of $\SI{13}{\tera\electronvolt}$ and integrated luminosity of $\SI{36.1}{\per\femto\barn}$. 

Events are required to satisfy the conditions $\met>\SI{250}{\giga\electronvolt}$, leading-$p_T>\SI{250}{\giga\electronvolt}$ and also
$|\eta|_{\mathrm{leading-jet }}<2.4$. In addition, a maximum of four jets with $p_T>\SI{30}{\giga\electronvolt}$ and $|\eta|<2.8$ are allowed, and the condition $\Delta\phi(\mathrm{jet},\boldsymbol p_T^{\rm{miss}})>0.4$ must be satisfied for each selected jet.
The analysis then uses ten different signal regions, which differ from each other by the choice of cut on $\met$: in particular, the weakest one is denoted (for the inclusive analysis) by IM1, and requires $\met>\SI{250}{\giga\electronvolt}$; while IM10 requires $\met>\SI{1000}{\giga\electronvolt}$.

We simulate events at parton level using \textsc{Madgraph5\_aMC@NLO}~\cite{Alwall:2014hca}, then apply the same cuts as Ref.~\cite{Aaboud:2017phn}. 
In models where a colored partner is produced at the LHC, monojet constraints will only apply if the colored partner decays promptly, i.e. within the beamline radius, $\ell_{\rm beam} = 2.5$ cm. Otherwise, if it enters the detector material, it will form an $R$-hadron within a very short timescale, roughly $\Lambda_{\mathrm{QCD}}^{-1}\sim\SI{e-24}{\second}$. We take this into account by considering the probability for each particle $\chi_2$ to decay with transverse decay length $\ell_T$ less than $d_{\rm beam}$ \cite{ElHedri:2017nny}:

\begin{equation}
\mathcal{P}(\ell_T<\ell_\text{beam})=1 - \exp\left(-\dfrac{\ell_\text{beam}}{\ell_T}\right)\, ,
\end{equation}
where $\ell_T=p_2^T(i)/(m_2\,\Gamma_{\chi_2})$ is the transverse distance traveled by $\chi_2$ in an event $i$. Each event is weighted by this probability in order to find the effective cross-section where $\chi_2$ decays promptly, before forming an $R$-hadron. We assume here that all the colored particles reaching the detecting stage hadronize.

In order to obtain a limit on the number $N_\text{NP}$ of new physics events, for both current and future luminosities, we apply a $\chi^2$ analysis with $95\%$ CL with unit efficiency and acceptance, according to~\cite{deSimone:2014pda}:
\begin{equation}
\chi^2=\frac{\left[N_\text{obs}-(N_\text{SM}+N_\text{NP})\right]^2}{N_\text{NP}+N_\text{SM}+\sigma_\text{SM}^2}\,,
\end{equation}
where the error on the SM background is assumed to be normally distributed.

\begin{figure}[t]
\begin{center}
\includegraphics[width=0.55\textwidth]{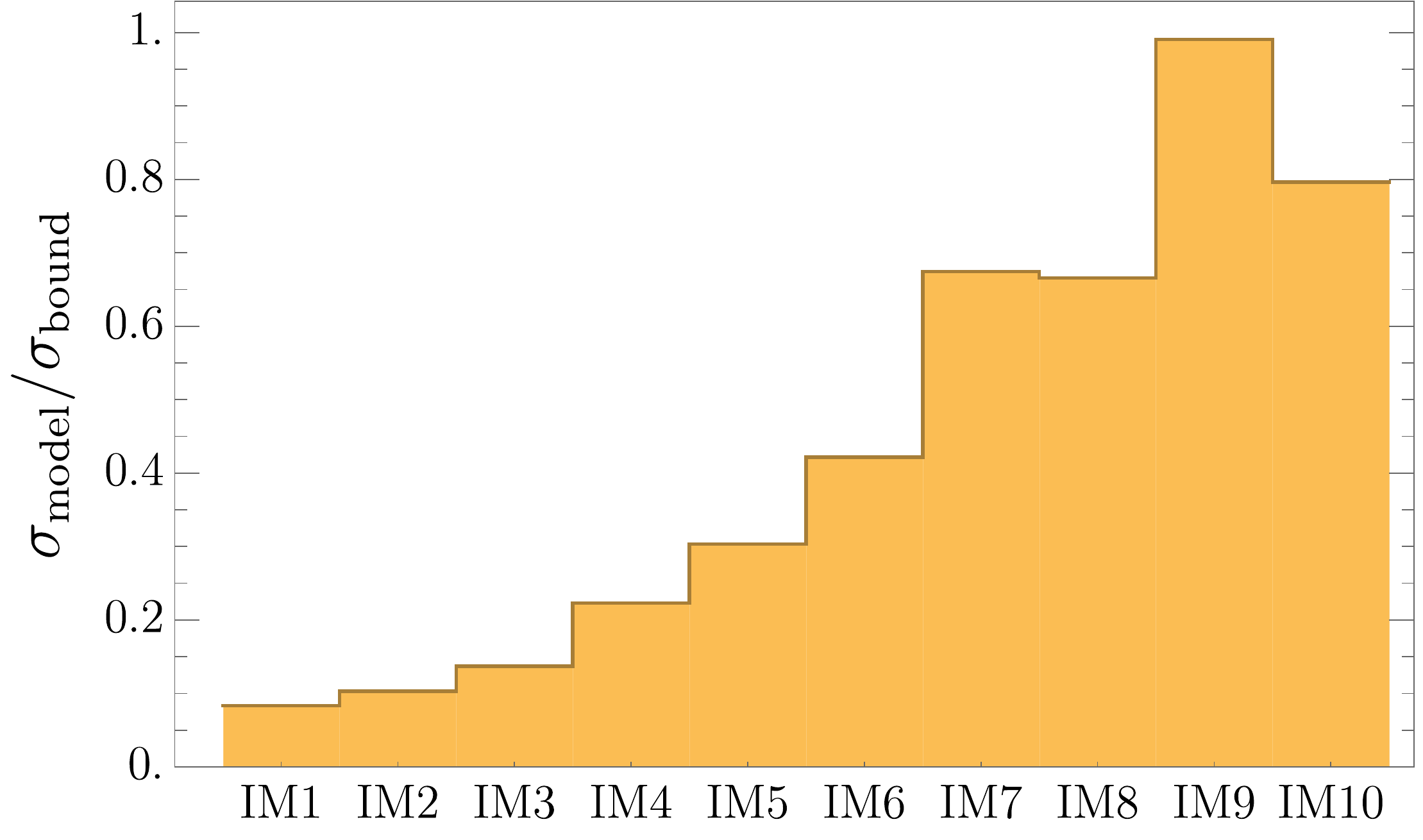}
\end{center}
\caption{\label{signal_regions} Ratio between the cross section from our model and the bound from Ref.~\cite{Aaboud:2017phn} in the case of $m_1=\SI{860}{\giga\electronvolt}$ as a function of the inclusive regions.}
\end{figure}

To find the strongest constraint, we consider the different signal regions from Ref.~\cite{Aaboud:2017phn}, differing by the cut on $\met$. For a given value of $m_1$, we use the ratio between our simulated cross section and the bound from the ATLAS paper and find that the strongest bound comes from IM9 (which requires $\met>\SI{900}{\giga\electronvolt}$) as can be seen in fig.~\ref{signal_regions}. It should be noted that changing the mass varies both the value of the cross section and the kinematic distribution of the particles, so that the results from fig.~\ref{signal_regions} cannot be trivially recast into a bound on the mass.  

For our optimal bin, the number of events in this signal region is:
\begin{equation}
N_\text{SM} = 464\pm34\quad,\quad N_\text{obs}=468\,.
\end{equation}
Then the cross section of new physics (NP) has to satisfy the constraint $\sigma_{NP}<\SI{2.3}{\femto\barn}$ for $\mathcal L=\SI{36.1}{\per\femto\barn}$.
Using this value and the procedure outlined earlier in this section, we find a lower bound on the mass of the DM of $\SI{860}{\giga\electronvolt}$ for $d_\chi \gtrsim 3\times 10^{-7}$. Full results are shown as the blue lines in Fig.~\ref{fig:chromo_comparison}. For smaller values of $d_\chi$, $\chi_2$ begins to travel into the detector and form $R$-hadrons before decaying, as discussed earlier in this section.

We extrapolate the monojet bound from Ref.~\cite{Aaboud:2017phn} to higher luminosity by considering the statistical and systematic uncertainties separately. The relative statistical error scales with the inverse square root of the number of events (and hence of the luminosity); on the other hand, it is generally not straightforward to predict how the relative systematic uncertainty will evolve with the luminosity. For this reason, we parametrize it in general as $\delta^{\mathrm{sys}}(\mathcal L_2)\equiv r\,\delta^{\mathrm{sys}}(\mathcal L_1)$. 
Using the published upper bound on the cross-section of new physics, $\sigma_{NP}$, at a luminosity $\mathcal L_1$, we can then estimate the corresponding upper bound on $\sigma_{NP}$ at a different luminosity $\mathcal L_2$ as:
\begin{equation}
\sigma_{NP}^{(\mathcal L_2)}(r)\leq\sigma_{NP}^{(\mathcal L_1)}\,\sqrt{r^2+\left(\frac{\mathcal L_1}{\mathcal L_2}-r^2\right)\frac{N_1}{\delta N_1^2}}\,.
\label{projection_sigma}
\end{equation}
We carry out this HL projection to 3000 fb$^{-1}$, in an optimistic scenario where systematic uncertainties have been cut to half the current values. In this case, we find a limit on $m_1$ from monojet of $\SI{1020}{\giga\electronvolt}$. Results are shown as the dashed blue curve in Fig.~\ref{fig:chromo_comparison}.

For reasons of completeness, we also considered the extreme cases in which the systematics will be unchanged with respect to their current value and the case in which they will be completely negligible, getting respectively the bounds $m_1>\SI{900}{\giga\electronvolt}$ and $m_1>\SI{1250}{\giga\electronvolt}$.

\subsection{Comparison between different searches}\label{sec:comparison}

\begin{figure}
\centering\includegraphics[width=0.8\textwidth]{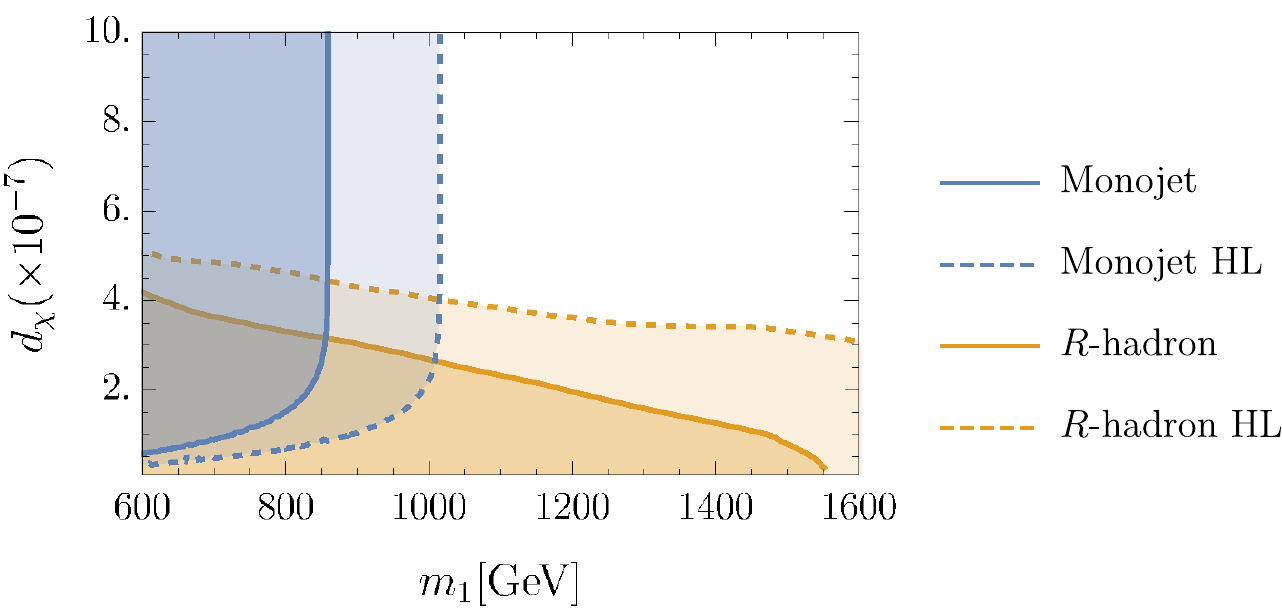}
\caption{Current (solid) and foreseen future (dashed) status of the parameter space as excluded by monojet and $R$-hadrons searches, in blue and orange, respectively.  The mass splitting $\Delta m$ is fixed, for given $m_1$, by the relic density as shown in fig.~\ref{relic:chromo}. The assumptions made regarding the scaling of the error at high luminosity can be found in the text.}\label{fig:chromo_comparison}
\end{figure}

The chromo-electric dipole dark matter model has been analyzed in the light of different LHC signals, namely monojet and $R$-hadrons. 

Since the different LHC signals are most effective in different regions of parameter space, it is important to understand the interplay between them. A first noteworthy feature is that the monojet analysis is insensitive to the value of the BSM coupling $d_\chi$ in most of the parameter space, but at some point this search becomes ineffective due to the fact that the colored partners are forming $R$-hadrons, rather than decaying to the DM. Since the observed events in this region of parameter space are the $R$-hadrons, the search for these states becomes the one giving the most stringent bound, as can be seen from fig.~\ref{fig:chromo_comparison}, where the result of the previous section are summarized. Note that the $R$-hadron results are shown here in terms of $m_1$, since relic density fixes $\Delta m$ for given $m_1$ and $d_\chi$.

The complementarity of the searches emerges from the fact that for $d_\chi\gtrsim\SI{3e-7}{}$ the most stringent bound is given by monojet searches, while for $d_\chi\lesssim\SI{3e-7}{}$ the $R$-hadron search gives the best result.

Furthermore, the different analyses are affected by different errors, meaning that increased luminosity has a distinct effect on each of them. This suggests that a high-luminosity projection might tell us which of these searches will become more interesting in the future. This as well is shown in fig.~\ref{fig:chromo_comparison}, where the role of higher luminosity in probing the parameter space is manifest.

As a side remark, we also checked the indirect detection limits by applying  bounds on the self-annihilation rate derived from cosmic-antiproton fluxes~\cite{Cuoco:2017iax} to our model.
An upper bound on $d_\chi$, as a function of $m_1$, is obtained from the upper bound on the annihilation cross section $\sigma(\chi_1\chi_1\rightarrow g g)$. The bounds were found to be very weak compared to those from collider searches, and in a region where the requirement $d_\chi \ll 1$ was not satisfied, e.g. the upper limit on $d_\chi$ was found to be $d_\chi\leq0.2$ for $m_1=\SI{1}{\tera\electronvolt}$ and $d_\chi\leq1$ for $m_1=\SI{5}{\tera\electronvolt}$.

\section{Conclusions}
\label{sec:conclusion}

In this paper, we have explored a remarkable possibility for DM phenomenology at the LHC: the combination of monojet and $R$-hadron searches. We performed our analysis using a simple effective operator of dark matter, as a case study giving rise to such a situation.

Since the cosmological abundance is dominated by QCD interactions, the coupling of the effective operator $d_\chi$ is not fixed by the relic density requirement, but it remains a free parameter. The only assumption we make is $d_\chi\ll1$, in order for the effective theory to be reliable. 

If $d_\chi$ is small enough, the chemical equilibrium can break down before the dark matter freeze-out. We analyzed such a situation and concluded that for the parameter space of interest for LHC searches ($d_\chi\gtrsim10^{-8}$) there is no need to take into account the breakdown of chemical equilibrium.

Our main analysis consisted of the combination  of monojet and $R$-hadrons searches, and found the regions of the $(m_1,d_\chi)$ parameter space excluded by current searches (see Figure \ref{fig:chromo_comparison}): while current monojet is able to exclude all points of the parameter space for $m_1\lesssim\SI{900}{\giga\electronvolt}$ for $d_\chi\gtrsim3\times10^{-7}$, current $R$-hadron results are instead able to constrain the parameter space  for larger masses, but smaller couplings. This complementarity is maintained in higher-luminosity projections. 

These results show once more the importance of finding complementary phenomenological signatures and the power of their combination in strengthening the reach of LHC searches for dark matter.

\acknowledgments
We would like to thank M.~Rinaldi, R.~Rattazzi and F.~Riva for insightful discussions.

\appendix

\section{Estimate of dijet constraints}
\label{app:dijet}

If a pair of $\chi_2$ are produced near rest, then rather than promptly decaying or forming $R$-hadrons, they can combine to form a QCD bound state analogous to gluinonium. 

The $\chi_2 \chi_2$ production rate is controlled only by QCD processes, and so should be model-independent. Therefore we use the production rate of gluinonia as calculated by Ref.~\cite{Kauth:2009ud} as an estimate of the production rate of $\chi_2 \chi_2$ bound states. 
The strongest constraints on this channel come from limits on the dijet resonance production cross-section. We use model-independent constraints from ATLAS \cite{Aaboud:2017yvp}, taking the conservative choice of assuming a narrow Gaussian width. We also use these limits to estimate the future high-luminosity constraints at 3000 fb$^{-1}$ using the same method as in previous sections. In order to evaluate the bounds, we have worked under the assumption that the fitting function for the dijet mass distribution used in \cite{Aaboud:2017yvp} is still suitable at higher luminosities and we have also considered the case in which the systematic uncertainties are unchanged in the projection.

\begin{figure}
\begin{center}
\includegraphics[scale=0.35]{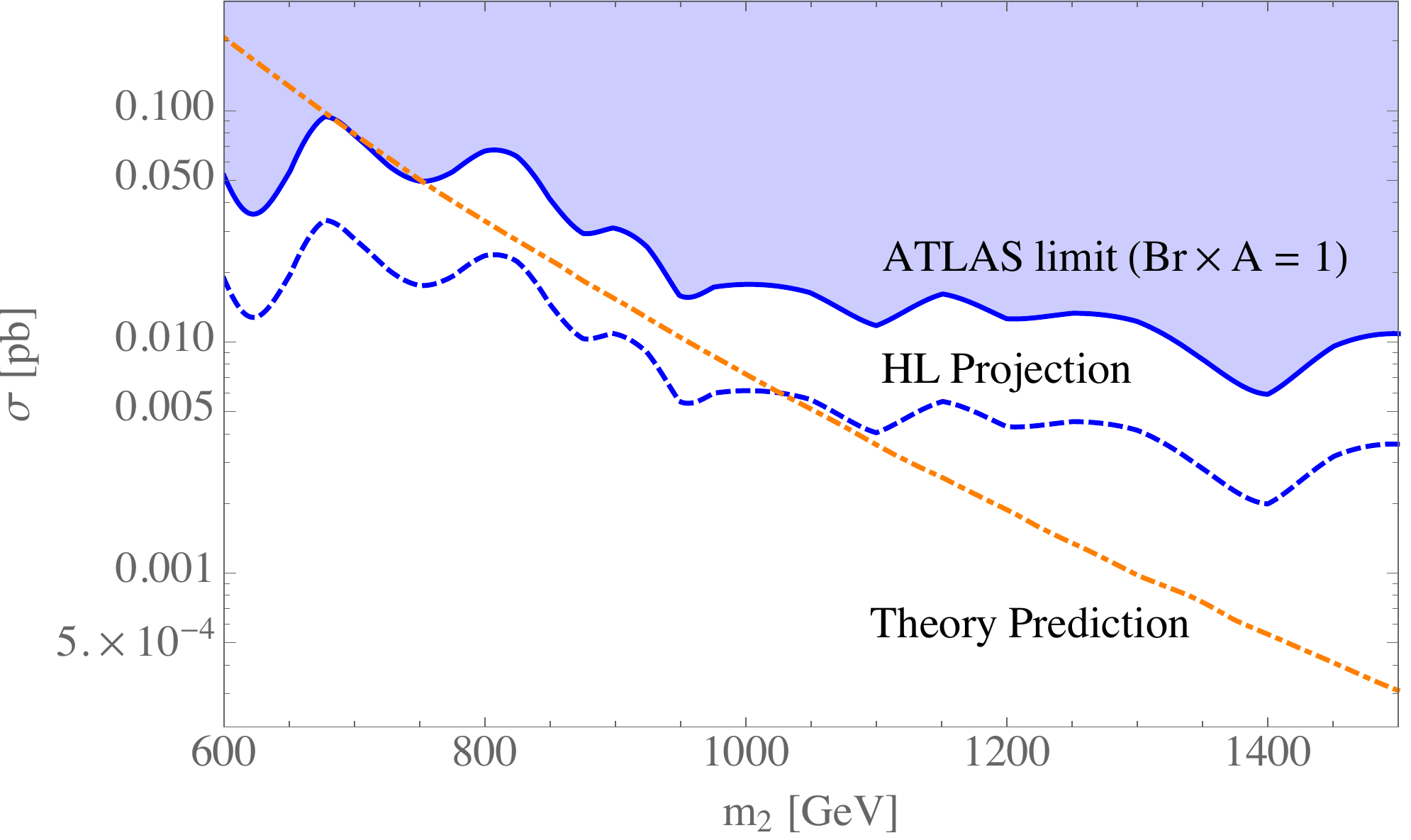}
\includegraphics[scale=0.35]{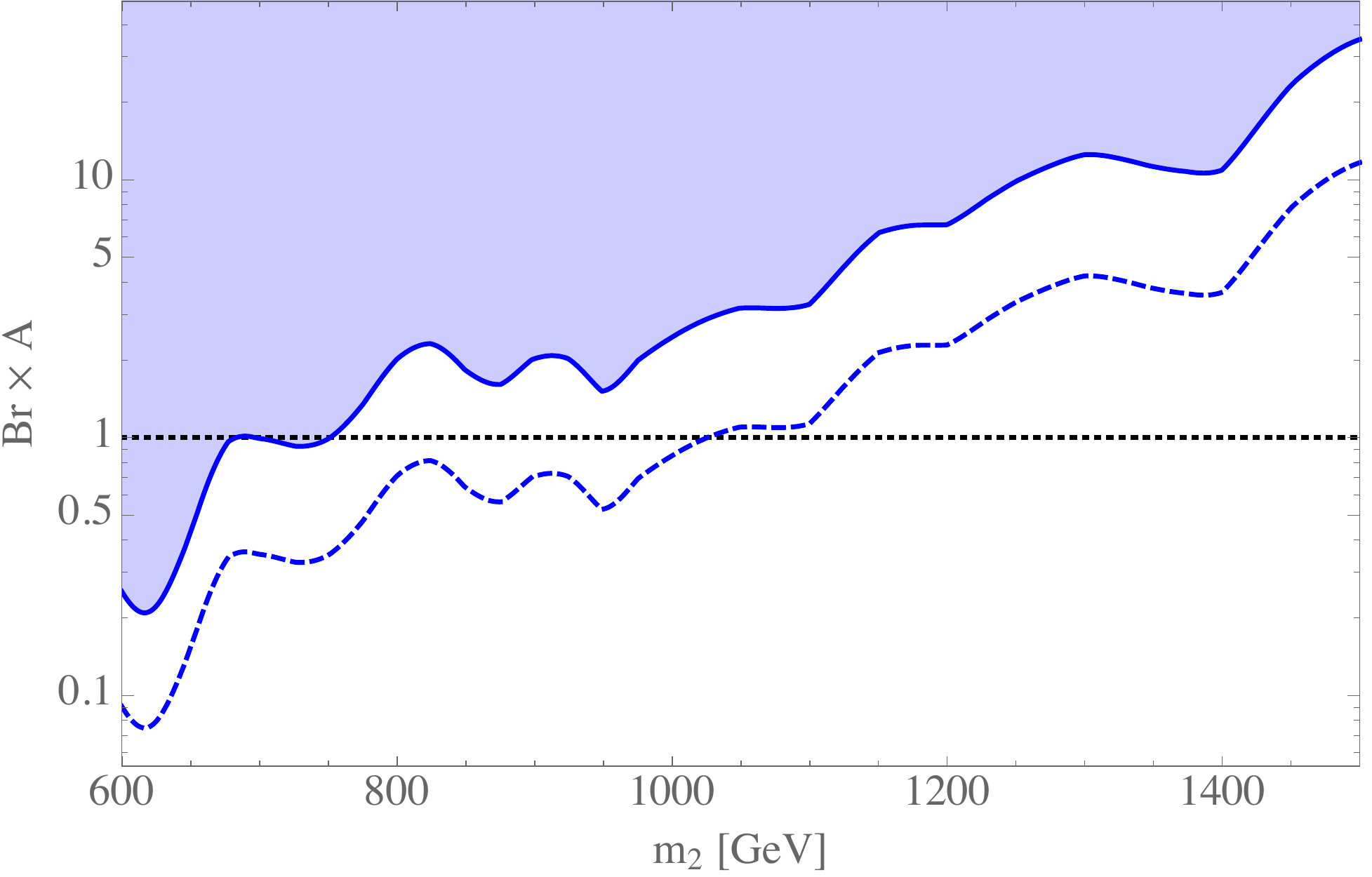}
\caption{\label{fig:dijet}\emph{left:} ATLAS excluded (shaded) and projected future limit (dashed) on dijet production cross-section assuming 100\% acceptance and branching ratio into dijets, along with theoretical production cross-section (dot-dashed). \emph{right}: Corresponding current excluded (shaded) and future limit (dashed) on the $Br \times A$. Maximum possible value $Br \times A = 1$ shown as dotted line to guide the eye. }
\end{center}
\end{figure}

These constraints are shown in fig.~\ref{fig:dijet} (left), along with the theoretical production rate from Ref.~\cite{Kauth:2009ud}. The limits assume that both the branching ratio to dijets ($Br$) and the acceptance ($A$) are 1. In reality $Br \times A <1$, and the limits are weakened proportionally. In fig.~\ref{fig:dijet} (right), these same results are visualized as an upper limit on $Br \times A$, above which the model is ruled out.

For values of $Br \times A$ greater than the maximum theoretical value of 1, the model is not currently constrained by dijets. For $m_2 \lesssim$ 650 GeV, dijet constraints rule out the model only if $Br \times A$  is greater than around 0.1 -- 1.

This constraint is conservative for two reasons: first, the production rate calculation is performed at 14 TeV while the constraints are at 13 TeV. The true 13 TeV production cross-section will be slightly smaller than shown; second, we have taken the dijet constraints assuming a narrow Gaussian width. A broader width weakens the constraints, as seen in Fig.~5 of Ref.~\cite{Aaboud:2017yvp}. In conclusion, while dijet searches do not strongly constrain the model at the moment, they may be an interesting channel to study with future data.

\section{Analytical expressions for the differential cross sections}
\label{app:sigma}

Here we list the analytical expressions for the differential cross sections, separately for each process. The expressions are simplified by using the quantities ${E}_1=\sqrt{m_1^2+p^2}$ and ${E}_2=\sqrt{m_2^2+p^2}$, which are the energies of the incoming particles $\chi_1$ and $\chi_2$. The expressions for the cross sections are referred to the center of mass frame (CM), so that $p$ and $\theta$ must be interpreted as the momentum of the incoming particles in the CM frame and the angle between incoming and outcoming momenta in the CM frame. The decay to massive quarks is also considered, here $m_q$ refers to the mass of the final quarks and the value for the scattering cross section refers to just one flavor.

\subsection{$\chi_1\chi_1\rightarrow \textnormal{gg}$}

The self-annihilation of the DM particles into gluons proceeds via $t$ and $u$ channel exchanges of $\chi_2$.

\begin{align}
\begin{split}
\frac{d \sigma v}{d \Omega}\bigg\rvert_\text{CM}=&\frac{d_{\chi }^4E_1^2}{2 \pi ^2 m_1^4 \left(-2 E_1 p   \cos \theta+E_1^2+m_2^2+p^2\right){}^2 \left(2 E_1 p   \cos\theta +E_1^2+m_2^2+p^2\right){}^2}\\
&\Big[\left.4 E_1^6 \left(m_2^2+2 p^2   \sin ^2\theta\right.\right)\\
&\,\,\,+2 E_1^4 \left(2 m_2^2 p^2 (3-4  \cos(2 \theta ) )+4 m_2^4+p^4  \sin ^2\theta  (-(7  \cos(2 \theta ) +1))\right)\\
&\,\,\, +E_1^2 \left(m_2^2 p^4 (7  \cos (4 \theta )-12  \cos(2 \theta ) +1)\right)\\
&\,\,\,+E_1^2 \left(4 m_2^4 p^2 (1-5  \cos (2 \theta ))+4 m_2^6+p^6  \sin (2 \theta )  \sin (4 \theta )\right)\\
&\,\,\, -4 p^2 \left(m_2^2+p^2\right){}^2 \left(p^2   \sin ^4\theta-m_2^2\right)\Big]
\end{split}
\end{align}

\subsection{$\chi_1\chi_2\rightarrow \textnormal{gg}$}

The co-annihilation into gluons proceeds via $s$-channel gluon interaction, $t$ and $u$ exchange of $\chi_2$ and via the quartic $\chi_1\chi_2$gg coupling.

\begin{align}
\begin{split}
\frac{d \sigma v}{d \Omega}\bigg\rvert_\text{CM}=&\frac{3 g_s^2 d_{\chi }^2}{512 \pi ^2 E_1 E_2 m_1^2 \left(E_2^2-p^2  \cos ^2\theta\right){}^2}\\
&\Big[-8 E_1^2 p^2 \left(E_2^2 (1-2  \cos (2 \theta ))+p^2  \cos ^2\theta\right)\\
&\,\,\, +8 E_1 E_2 p^2 \left(E_2^2 \left(4-5\cos ^2\theta\right)+p^2   \cos ^2\theta \left(4-3  \cos ^2\theta\right)\right)\\
&\,\,\, +8 E_1^3 E_2 \left(E_2^2-p^2 \cos ^2\theta \right)\\
&\,\,\,+ E_2^2 p^2 \left(4 m_1m_2(1-\cos(2\theta))+p^2 (3  \cos(4 \theta ) -11)\right)\\
&\,\,\,+ p^4 m_1m_2 ( \cos (4 \theta )-1)+p^6 ( \cos(4 \theta ) +7) \cos ^2\theta\Big]
\end{split}
\end{align}

\subsection{$\chi_1\chi_2\rightarrow \textnormal{qq}$}

The co-annihilation into quarks proceeds via $s$-channel gluon interaction.

\begin{align}
\begin{split}
\frac{d \sigma v}{d \Omega}\bigg\rvert_\text{CM}=&\frac{g_s^2 d_{\chi }^2 \sqrt{\left(E_1+E_2\right){}^2-4 m_q^2} }{64 \pi ^2 E_1 E_2 \left(E_1+E_2\right){}^3 m_1^2}\\
 &
\Big[p^2 \left(2 m_q^2 \cos(2 \theta ) -\left(E_1+E_2\right){}^2  \cos ^2\theta\right)(E_1 E_2 -m_1m_2)\left(\left(E_1+E_2\right){}^2+2 m_q^2\right)]
\end{split}
\end{align}

\subsection{$\chi_2\chi_2\rightarrow \textnormal{gg}$ }

The annihilation of the partners into gluons proceeds via five different channels: $t$ and $u$ exchange of $\chi_1$, whose amplitude will be proportional to $d_\chi^2$, $s$-channel gluon interaction, and $t$ and $u$ exchange of $\chi_2$, whose amplitude will be proportional to $g_s^2$.

Hence the total cross section will be the sum of three terms proportional to $g_s^4$, $g_s^2 d_\chi^2$ and $d_\chi^4$.

\begin{align}
\begin{split}
\frac{d \sigma v}{d \Omega}\bigg\rvert_\text{CM}=&-\frac{9 g_s^4}{16384 \pi ^2 E_2^4 \left(E_2^2-p^2  \cos ^2\theta\right){}^2}\\
& \Big[E_2^2 p^4 (5  \cos(4 \theta ) -12 \cos (2 \theta )+31)+4 E_2^4 p^2 (5 \cos (2 \theta )-7)\\
&\,\,\,-24 E_2^6+p^6 ( \cos (4 \theta )-4 \cos (2 \theta ) +11) \cos ^2\theta\Big]\\
&\\
+&\frac{3 g_s^2 d_{\chi }^2}{512 \pi ^2 m_1^2  (E_2^2-p^2   \cos^2 \theta) (-2 E_2 p   \cos\theta +E_2^2+m_1^2+p^2) (2 E_2 p   \cos \theta+E_2^2+m_1^2+p^2)}\\
& \Big[4 E_2^4 (m_1 m_2+2 p^2   \sin ^2\theta)+2 m_1 p^2 (-2 m_1^2 m_2  \cos (2 \theta )+p^2 m_2 ( \cos(4 \theta ) +1)\\
&\,\,\, +m_1 p^2  \sin ^2\theta  (3  \cos (2 \theta )+1))-2 E_2^2 (4 m_1 p^2 m_2  \cos (2 \theta )-2 m_1^3 m_2-4 m_1^2 p^2  \sin ^2 \theta \\
&\,\,\, +p^4  \sin ^2 \theta  (5 \cos(2 \theta )  +3))+p^6  \sin ^2 \theta (6  \cos(2 \theta ) - \cos(4 \theta ) +3)\Big]\\
&\\
+&\frac{d_{\chi }^4 E_2^2}{64 \pi ^2 m_1^4 (-2 E_2 p   \cos\theta +E_2^2+m_1^2+p^2){}^2 (2 E_2 p  \cos \theta+E_2^2+m_1^2+p^2){}^2}\\
& \Big[2 E_2^6 (8 m_1^2+9 p^2  \sin ^2\theta )+E_2^4 (m_1^2 p^2 (27-43  \cos(2 \theta ) )+32 m_1^4+2 p^4  \sin ^2\theta (3-14 \cos (2 \theta )))\\
&\,\,\,+E_2^2 (m_1^2 p^4 (21 \cos (4 \theta ) -34  \cos (2 \theta )-3)\\
&\,\,\, -m_1^4 p^2 (59  \cos (2 \theta )+5)+16 m_1^6+2 p^6  \sin ^2\theta (4  \cos (4 \theta )-6 (\cos (2 \theta))-3))\\
&\,\,\, +p^2 (m_1^2+p^2){}^2 (m_1^2 (7  \cos(2 \theta ) +9)+2 p^2   \sin ^2\theta (4  \cos(2 \theta ) +3))\Big]
\end{split}
\end{align}

\subsection{$\chi_2\chi_2\rightarrow \textnormal{qq}$}

The annihilation of the partners into quarks proceeds via purely QCD $s$-channel gluon interactions.

\begin{equation}
\frac{d \sigma v}{d \Omega}\bigg\rvert_\text{CM}=\frac{3 g_s^4 \sqrt{E_2^2-m_q^2} \left(E_2^2 \left(m_q^2+p^2  \cos ^2\theta-p^2\right)+2 E_2^4-m_q^2 p^2   \cos ^2\theta\right)}{2048 \pi ^2 E_2^7}
\end{equation}

\bibliographystyle{JHEP} \bibliography{bibliography} 
\end{document}